\documentstyle[preprint,tighten,aps,floats,epsfig,colordvi,psfig]{revtex}

\begin{document}
\draft
\vskip 2cm

\title{The Lattice Free Energy of QCD with Clover Fermions, up to Three-Loops}

\author{A. Athenodorou$^{\rm a}$, H. Panagopoulos$^{\rm b}$ and
  A. Tsapalis$^{\rm c}$\\
\phantom{a}}

\address{$^{\rm a}$Rudolf Peierls Centre for Theoretical Physics,
  University of Oxford,\\
1 Keble Road, Oxford, OX1 3NP, U.K. \\
$^{\rm b}$Department of Physics, University of Cyprus,\\
P.O. Box 20537, Nicosia CY-1678, Cyprus \\
$^{\rm c}$Institute of Accelerating Systems and Applications,\\
University of Athens, Greece\\
\medskip
{\it email: }{\tt a.Athenodorou1@physics.ox.ac.uk, haris@ucy.ac.cy,
  a.tsapalis@iasa.gr}}
\vskip 3mm

\date{\today}

\maketitle

\begin{abstract}

We calculate the perturbative value of the free energy in Lattice QCD, 
up to three loops. Our calculation is performed using Wilson gluons
and the Sheikholeslami - Wolhert
(clover) improved  action for fermions.

The free energy is directly related to the average plaquette.
To carry out the calculation, we compute all relevant Feynman diagrams 
up to 3 loops, using a set of automated procedures in Mathematica;
numerical evaluation of the resulting loop integrals is performed on
finite lattice, with subsequent extrapolation to infinite size. 

The results are presented as a function of the fermion mass $m$, for
any $SU(N_c)$ gauge group, and for an arbitrary number of fermion flavors.
In order to enable independent comparisons,
we also provide the results on a {\it per diagram} basis, 
for a specific mass value.   



\end{abstract}

\newpage


\section{Formulation of the Problem}

In this work we calculate the free energy of QCD on the lattice, 
up to three loops in perturbation theory. 
We employ Wilson gluons and the ${\cal O}(a)$ improved Sheikholeslami-Wohlert 
(clover)~\cite{SW} action for fermions. 
The purpose of this action is to reduce finite lattice spacing effects, 
leading to a faster approach to the continuum. Dynamical
simulations employing the clover action are currently in progress by the
CP-PACS/JLQCD~\cite{CPPACS} and UKQCDSF~\cite{UKQCDSF} collaborations and therefore perturbative studies of
properties of the QCD action with clover quarks are worthy of being
undertaken. The free energy, in the simpler case of 
Wilson fermions, was studied in~\cite{AFP}. 

The free energy in QCD on the lattice can be related to the average plaquette. 
The results find several applications, for example: 
a) In improved scaling schemes, using an appropriately defined
effective coupling which depends on the average plaquette (see, e. g.,
\cite{Parisi,L-M}),
b) In long standing efforts, starting with~\cite{GR}, to determine the
value of the gluon condensate,  
c) In studies of the interquark potential~\cite{BB}, and 
d) As a test of perturbation theory, at its limits of applicability.

Indeed, regarding point (d) above, the plaquette expectation value is
a prototype for additive renormalization of a composite, dimensionful
operator. The vacuum diagrams contributing to such a calculation are
power divergent in the lattice spacing and may well dominate over any
nonperturbative signal in a numerical simulation.

Starting from the Wilson formulation of QCD on the lattice, 
with the addition of the clover (SW) fermion term, the action reads in
standard notation: 
\begin{eqnarray}
S_L &\equiv& S_G + S_F\,,
\nonumber \\
S_G &=&  {1\over g^2} \sum_{x,\,\mu,\,\nu}
{\rm Tr}\left[ 1 - U_{\mu,\,\nu}(x) \right],
\nonumber \\
S_F &=& \sum_{f}\sum_{x} (4r+m_B)\bar{\psi}_{f}(x)\psi_f(x) 
\nonumber \\
&&-{1\over 2}\sum_{f}\sum_{x,\,\mu}
\left[ 
\bar{\psi}_{f}(x)\left( r - \gamma_\mu\right)
U_{\mu}(x)\psi_f(x+\hat{\mu})+
\bar{\psi}_f(x+\hat{\mu})\left( r + \gamma_\mu\right)
U_{\mu}(x)^\dagger
\psi_{f}(x)\right]\nonumber \\
&&+ {i\over 4}\,c_{\rm SW}\,\sum_{f}\sum_{x,\,\mu,\,\nu} \bar{\psi}_{f}(x)
\sigma_{\mu\nu} {\hat F}_{\mu\nu}(x) \psi_f(x)
\label{latact}
\end{eqnarray}
\begin{equation}
{\rm where:}\qquad {\hat F}_{\mu\nu} \equiv {1\over8}\,
(Q_{\mu\nu} - Q_{\nu\mu}), \qquad 
Q_{\mu\nu} = U_{\mu,\,\nu} + U_{\nu,\,{-}\mu} + U_{{-}\mu,\,{-}\nu} + U_{{-}\nu,\,\mu}
\end{equation}

Here $U_{\mu,\,\nu}(x)$ is the usual product of $SU(N_c)$ link variables
$U_{\mu}(x)$ along the perimeter of a plaquette in the $\mu$-$\nu$
directions, originating at $x$;
$g$ denotes the bare coupling constant; $r$ is the Wilson parameter,
which will be assigned its standard value $r=1$;
$f$ is a flavor index; $\sigma_{\mu\nu} =
(i/2) [\gamma_\mu,\,\gamma_\nu]$. 
Powers of the lattice spacing $a$ have been omitted and 
may be directly reinserted by dimensional counting.

The clover coefficient $c_{\rm SW}$ is a free parameter for the
purposes of the present calculation and our results will be presented
as a polynomial in $c_{\rm SW}$\,, with coefficients which we
compute. Preferred values for $c_{\rm SW}$ have been suggested by both
perturbative (1-loop)~\cite{SW} and non-perturbative~\cite{Luscher1996} studies.

We use the standard covariant gauge-fixing term~\cite{KW}; in terms of
the vector field $Q_\mu(x)$ $\left[U_{\mu}(x)= \exp(i\,g\,Q_\mu(x))\right]$, it
reads:
\begin{equation}
S_{\rm gf} = \lambda_0 \sum_{\mu , \nu} \sum_{x}
\hbox{Tr} \, \Delta^-_{\mu} Q_{\mu}(x) \Delta^-_{\nu} Q_{\nu}(x), \qquad
\Delta^-_{\mu} Q_{\nu}(x) \equiv Q_{\nu}(x - {\hat \mu}) - Q_{\nu}(x)
\end{equation}

Having to compute a gauge invariant quantity, we chose to work
in the Feynman gauge, $\lambda_0 = 1$. 
Covariant gauge fixing produces the following
action for the ghost fields $\omega$ and $\overline\omega$

\begin{eqnarray}
&\displaystyle S_{\rm gh} = 2 \sum_{x,\mu} \hbox{Tr} \,
\biggl\{ \Bigl(\Delta^+_{\mu}\omega(x)\Bigr)^{\dagger} \Bigl( &\Delta^+_{\mu}\omega(x) +
i g_0 \left[Q_{\mu}(x),
\omega(x)\right] + \frac{i\,g_0}{2}
\,\left[Q_{\mu}(x), \Delta^+_{\mu}\omega(x) \right] \nonumber\\
 & & - \frac{g_0^2}{12}
\,\left[Q_{\mu}(x), \left[ Q_{\mu}(x),
\Delta^+_{\mu}\omega(x)\right]\right]\nonumber\\
&&  - \frac{g_0^4}{720}
\,\left[Q_{\mu}(x), \left[Q_{\mu}(x), \left[Q_{\mu}(x), \left[ Q_{\mu}(x),
\Delta^+_{\mu}\omega(x)\right]\right]\right]\right]
+ \cdots \Bigr)\biggr\} ,\nonumber\\
&\Delta^+_{\mu}\omega(x) \equiv \omega(x + {\hat \mu}) - \omega(x)&
\end{eqnarray}

Finally the change of integration variables from links to vector
fields yields a jacobian that can be rewritten as 
the usual measure term $S_m$ in the action:

\begin{equation}
S_{\rm m} = \sum_{x,\mu} \biggl\{ \frac{N_c\, g_0^2}{12}\, \hbox{Tr} \,
\bigl\{ \bigl(Q_\mu(x)\bigr)^2 \bigr\} + \frac{N_c\, g_0^4}{1440} \,\hbox{Tr} \,
\bigl\{ \bigl(Q_\mu(x)\bigr)^4 \bigr\} + \frac{g_0^4}{480} \,\Bigl(\hbox{Tr} \,
\bigl\{ \bigl(Q_\mu(x)\bigr)^2 \bigr\}\Bigr)^2 + \cdots \biggr\}
\end{equation}

In $S_{\rm gh}$ and $S_{\rm m}$ we have written out only
terms relevant to our computation.
The full action is: 

\begin{equation}
S = S_L + S_{\rm gf} + S_{\rm gh} + S_{\rm m}
\end{equation}

The average value of the action density, $S/V$, is directly related to
the average plaquette. For the gluon part we have: 
\begin{equation}
\langle S_G/V \rangle = 6 \,\beta\, E_G\,,\qquad
E_G\equiv 1 - {1\over N_c} \hbox{Tr}\langle U_{\mu,\,\nu}(x)\rangle,
\qquad \beta = {2N_c/g^2}
\end{equation} 
As for $\langle S_F/V\rangle$, it is trivial in any action which is
bilinear in the fermion fields, and leads to:
\begin{equation}
\langle S_F/V\rangle = - 4 N_c N_f
\label{ef}
\end{equation}
($N_f$\,: number of fermion flavors).

We will calculate $E_G$ in perturbation theory:
\begin{equation}
E_G = c_1 \; g^2 + c_2 \; g^4 + c_3 \; g^6 + \cdots
\label{expansion}
\end{equation}

The $n$-loop coefficient can be written as $c_n = c^G_n + c^F_n$ where
$c^G_n$ is the contribution of diagrams without fermion loops and
$c_n^F$ comes from diagrams containing fermions. The coefficients
$c^G_n$ have been known for some time up to 3 
loops~\cite{ACFP} (also in 3 dimensions~\cite{PST}, where they are
applied to ``Magnetostatic'' QCD~\cite{Hietanen} and to dimensionally
reduced QCD~\cite{Braaten,Linde}). Independent estimates of higher loop
coefficients have also been obtained using stochastic perturbation
theory~\cite{DiRenzo}. The fermionic coefficients $c_n^F$ are known to
2 loops for overlap fermions~\cite{AP} and up to 3
loops for Wilson fermions~\cite{AFP}; in the present work we extend
this computation to the clover action.

The calculation of $c_n$ proceeds most conveniently by computing first the free energy $-(\ln Z)/V$, where $Z$ is the full partition function:
\begin{equation}
Z \equiv \int [{\cal D}U {\cal D}\bar\psi_i {\cal D}\psi_i] \exp(-S) 
\label{Z}
\end{equation}
Then, $E_G$ is extracted through
\begin{equation}
E_G = - {1 \over 6}\, {\partial \over {\partial \beta}}\, \left( {\ln Z \over V} \right) 
\label{e}
\end{equation}
In particular, the perturbative expansion of $(\ln Z)/V$ :
\begin{equation}
(\ln Z)/V = d_0 -{3 (N_c^2{-}1)\over 2}\,\ln\beta + {d_1\over\beta} + {d_2\over\beta^2} + \cdots
\end{equation}
leads immediately to the relations:
\begin{equation}
c_2 = d_1/(24N_c^2), \qquad
c_3 = d_2/(24N_c^3)
\end{equation}

\section{Calculation and Results}

A Total of 62 Feynman diagrams contribute to the present calculation,
 up to three loops. The first 36 diagrams are totally gluonic, and the 
others have both gluon and fermion contributions; these are shown
in Appendix A. The involved algebra of lattice 
perturbation theory was carried out using our computer
package in Mathematica. The value for each diagram is computed numerically for a
sequence of finite lattices, with typical size $L \le 36$.

Certain diagrams must be grouped into infrared-finite
sets, before extrapolating their values to infinite lattice size
(diagrams 11+12+13, 22 through 36, 43+53, 44+52+58, 46+56, 51+57,
55+60, 61+62). Extrapolation leads to a (small) systematic error, which 
is estimated quite accurately; a consise description of the procedure
is provided in Ref.~\cite{PST}.

Diagrams in the shape of a triangular pyramid (18, 19, 20, 49, 50) are the most CPU
demanding, since integration over the 3 loop momenta cannot be
factorized; these diagrams were necessarily evaluated for smaller $L$,
but fortunately $L\sim 16$ was already sufficient for a very stable
extrapolation in these cases. Diagram 40 vanishes identically by color antisymmetry.

The pure gluon contributions are already known~\cite{ACFP}:
\begin{eqnarray}
 c_1^G &=& {{N_c^2 {-} 1} \over 8 \;N_c} , \\
 c_2^G &=& \left( N_c^2 {-} 1 \right) \left(0.0051069297 -
           {1 \over {128 \; N_c^2}}\right) , \nonumber \\
 c_3^G &=& \left( N_c^2 {-} 1 \right) \left( {0.0023152583(50) \over N_c^3}
- {0.002265487(17) \over N_c} + 0.000794223(19) \; N_c \right)  \nonumber 
\end{eqnarray}

Fermion contributions take the form:
\begin{eqnarray}
 c_1^F &=& 0  \; ,\nonumber \\
 c_2^F &=&(N_c^2 {-} 1) h_2 \; {{ N_f} \over N_c} , \\
 c_3^F &=& \left(N_c^2 {-} 1\right)
          \left( h_{30} \; N_f + h_{31} \; {N_f \over N_c^2} + 
           h_{32} \; {N_f^2 \over N_c}\right)  \nonumber
\end{eqnarray}

The coefficients $h_2, h_{30}, h_{31}, h_{32}$ depend polynomially on
the clover parameter $c_{\rm SW}$:
\begin{equation}
 h_2 {=} h_2^{(0)} {+} h_2^{(1)} \, c_{\rm SW} {+} h_2^{(2)} \, c_{\rm SW}^2 
\end{equation}

$$ h_{3i} {=} h_{3i}^{(0)} {+} h_{3i}^{(1)} \, c_{\rm SW} {+}
 h_{3i}^{(2)} 
 \, c_{\rm SW}^2 {+} h_{3i}^{(3)} \, c_{\rm SW}^3 {+} h_{3i}^{(4)} \,
 c_{\rm SW}^4 $$

We have calculated $h_2^{(j)}$, $h_{3i}^{(j)}$ 
for typical values of the Lagrangian (unrenormalized) fermion mass
parameter $m$, which is connected to the familiar
hopping expansion parameter $\kappa=1/(2m+8)$. Our results are listed in Appendix B. A complete {\it
  per diagram} breakdown of the results would be far too lengthy to
present; instead, for potential comparisons, we provide in Appendix B
a breakdown only for a particular value of $m$.

In Figs. 1 and 2 we present the dependence of $c_2$ and $c_3$\,,
respectively, on $m$, for three typical values of $c_{\rm SW}$.

\begin{center}
\epsfig{file=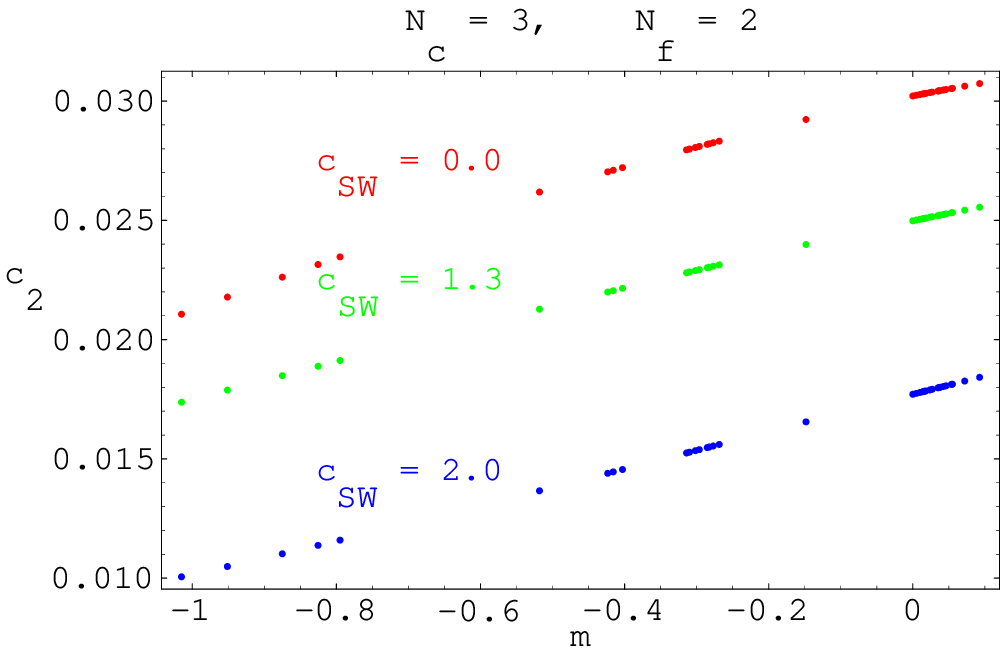,height=10truecm}\\
{\footnotesize Fig. 1. The dependence of $c_2$ on the fermion mass $m$,
  for some standard values of $c_{\rm SW}$\,. $N_c=3$, $N_f=2$.}
\end{center}

\begin{center}
\epsfig{file=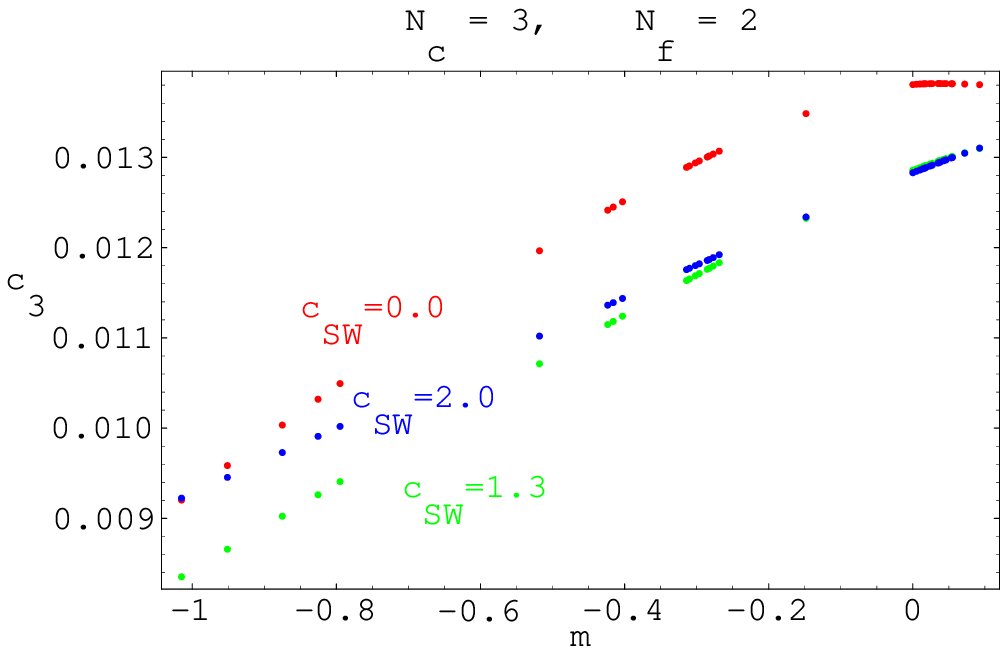,height=10truecm}\\
{\footnotesize Fig. 2. The dependence of $c_3$ on the fermion mass $m$,
  for some standard values of $c_{\rm SW}$\,. $N_c=3$, $N_f=2$.}
\end{center}

We list below some examples of values for $E_G$\,, setting $N_c=3$. 
For $N_f=0$ we have:
\begin{equation}
E_G = (1/3)\, g^2 + 0.0339109931(3) \, g^4
+ 0.0137063(2) \, g^6
\end{equation}
For two degenerate flavors ($N_f=2$) and
$m=-0.518106$ (corresponding to $\kappa = (8+2m)^{-1} = 0.1436$):
\begin{equation}
\begin{array}{llll}
c_{\rm SW} = 0.0 :\ &E_G = (1/3)\, g^2 &+ 0.026185200(3) g^4 &+ 0.0119649(3) g^6, \\[0.5ex]
c_{\rm SW} = 2.0 :\ &E_G = (1/3)\, g^2 &+ 0.013663456(3) g^4 &+ 0.0110200(13) g^6
\end{array}
\label{m-0.518106}
\end{equation}
For $N_f=2$ and $m=0.038$:
\begin{equation}
\begin{array}{llll}
c_{\rm SW} = 0.0 :\  &E_G = (1/3)\, g^2 &+ 0.030438866(3) \ g^4 &+ 0.0138181(2) \ g^6, \\[0.5ex]
c_{\rm SW} = 1.3 :\  &E_G = (1/3)\, g^2 &+ 0.025219798(9) \ g^4 &+ 0.0129659(5) \ g^6, \\[0.5ex]
c_{\rm SW} = 2.0 :\  &E_G = (1/3)\, g^2 &+ 0.01800170(1)  \ g^4 &+ 0.012948(1) \ g^6
\end{array}
\label{m0.038}
\end{equation}

It is seen that the 3-loop coefficients are quite pronounced for
typical values of $g$ used in numerical simulations. For convenience,
the behaviour of $E_G$ versus $\beta$ is also presented in Fig. 3, for
the same parameter values as in Eqs.~(\ref{m-0.518106},\ref{m0.038}).

\begin{center}
\epsfig{file=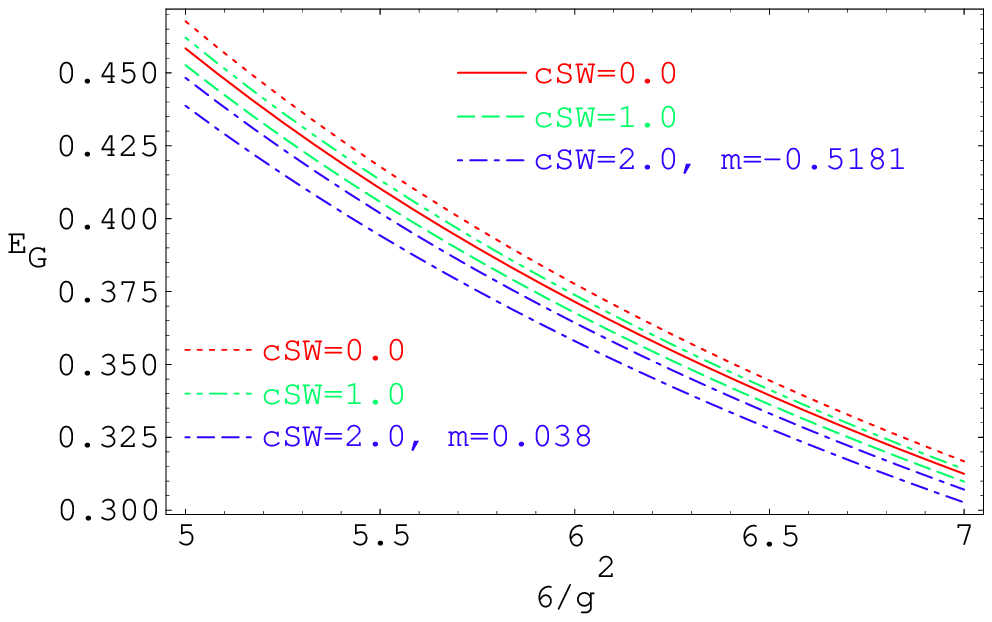,height=10truecm}\\
{\footnotesize Fig. 3. $E_G$ as a function
  of $\beta$, for  
$N_c=3$, $N_f=2$, and specific mass values.}
\end{center}

The detailed results, tabulated in Appendix B for arbitrary values of
$N_c$, $N_f$, $c_{\rm SW}$, show a very smooth behaviour as a function
of $m$; consequently, one is able to reconstruct $E_G$ also for
arbitrary values of $m$ by naive interpolation, to excellent
precision.


\eject
\appendix
\section{}
\label{appa}
\begin{center}
\epsfig{width=7.5truecm,file=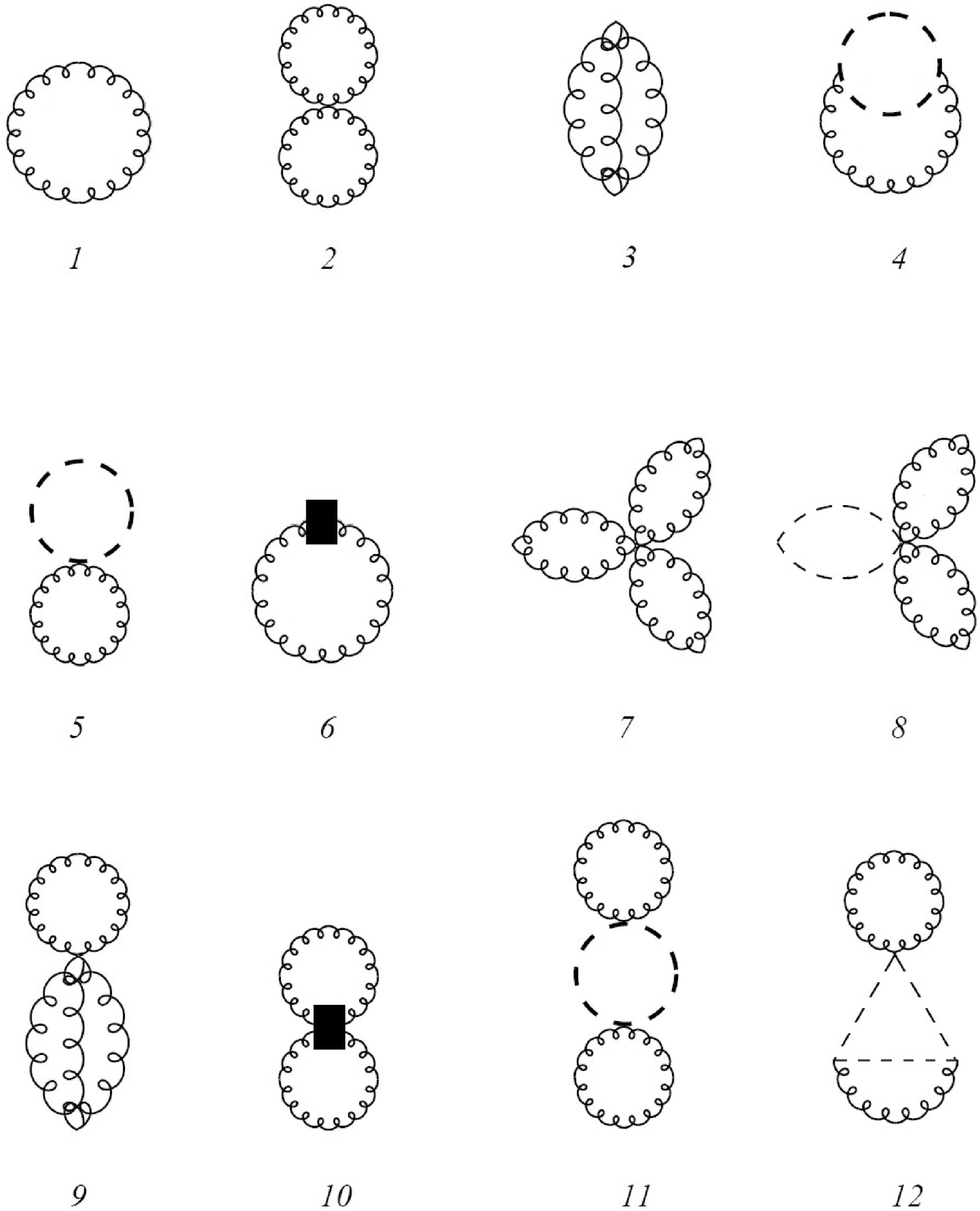}\hskip0.2truecm
\epsfig{width=7.5truecm,file=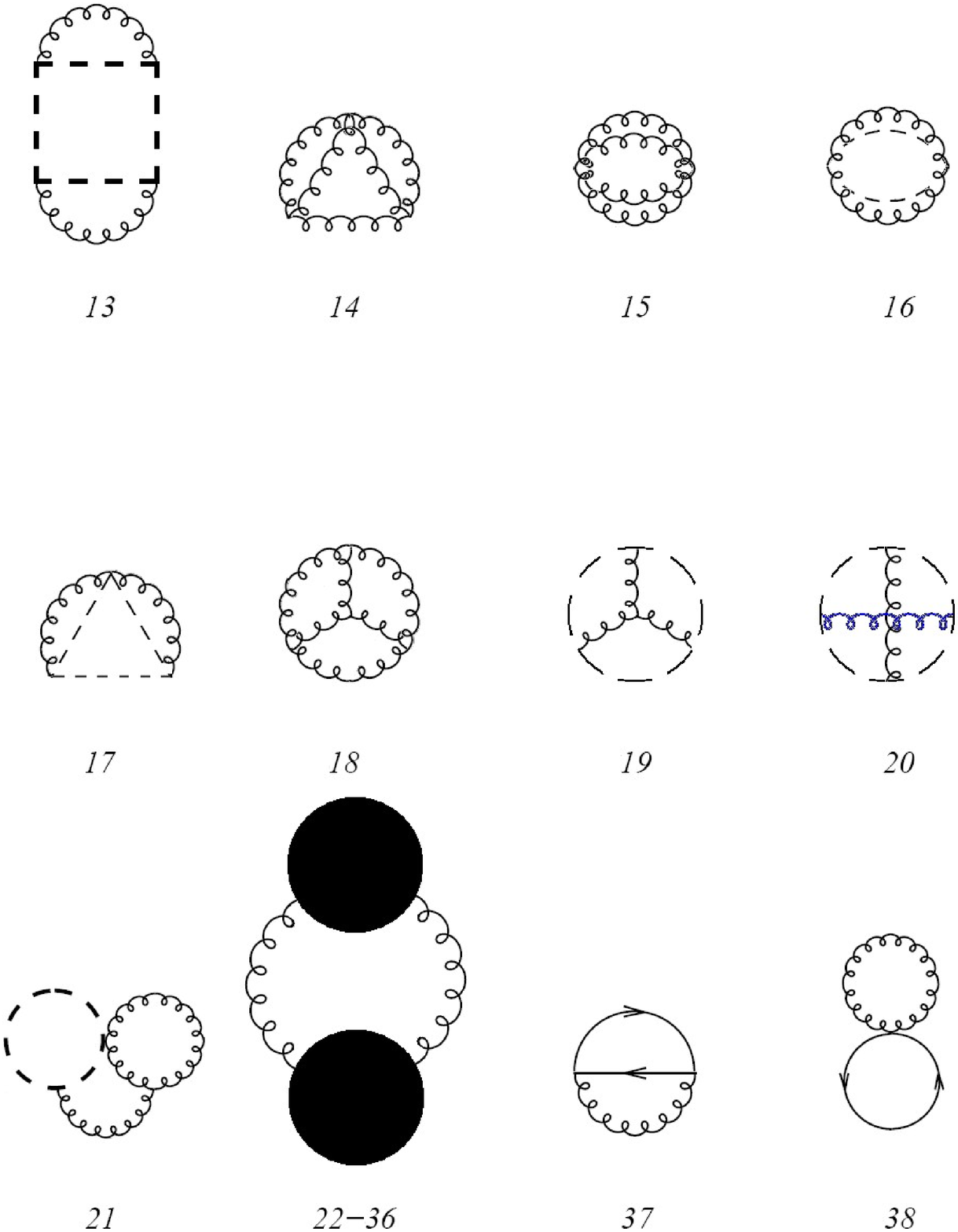}
\epsfig{width=7.5truecm,file=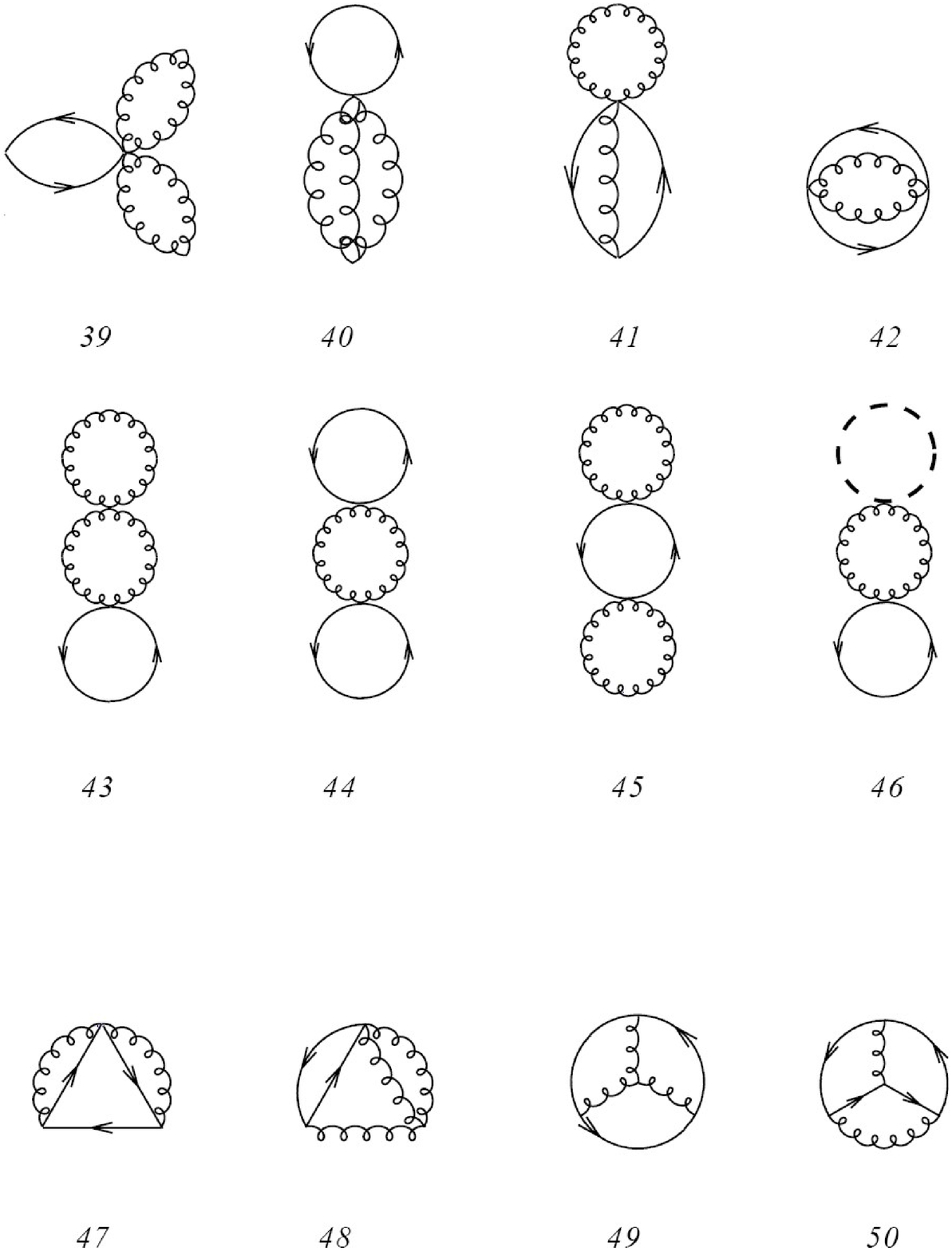}\hskip0.2truecm
\epsfig{width=7.5truecm,file=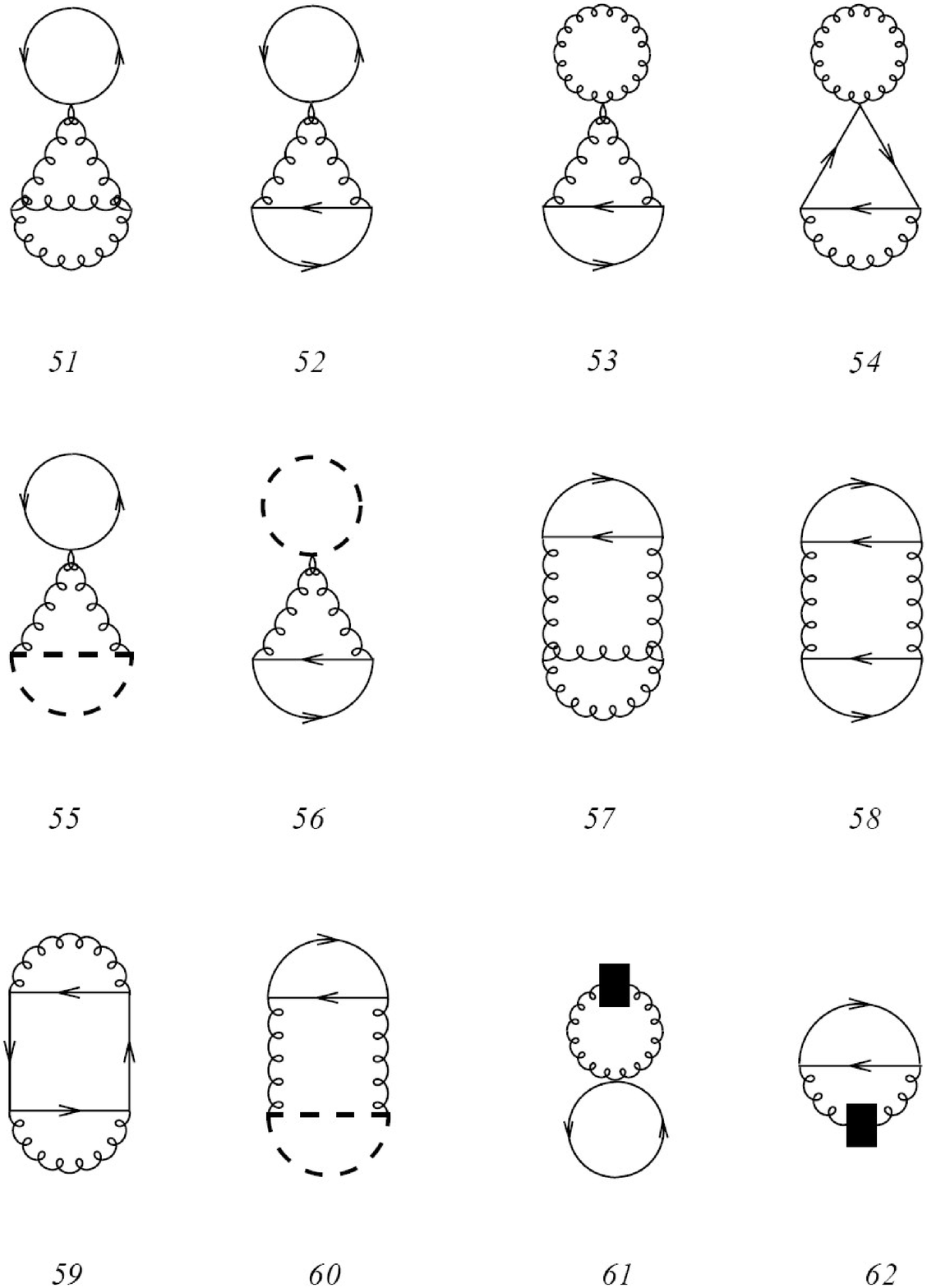}\\
{\footnotesize Fig. 4. Feynman diagrams contributing to the free
  energy, at one, two, and three loops.}
\end{center}

Figure 4 depicts all diagrams contributing to the free energy at 1
loop (diagram 1), 2 loops (2-6, 37-38), and 3 loops (7-36, 39-62). Solid (curly, dashed)
 lines represent fermions (gluons, ghosts), and the filled square is the contribution 
from the measure part of the action. The filled circle, corresponding
to the non-fermionic part of the 1-loop gluon self-energy, is given in
Figure 5.
\begin{center}
\epsfig{width=6truecm,file=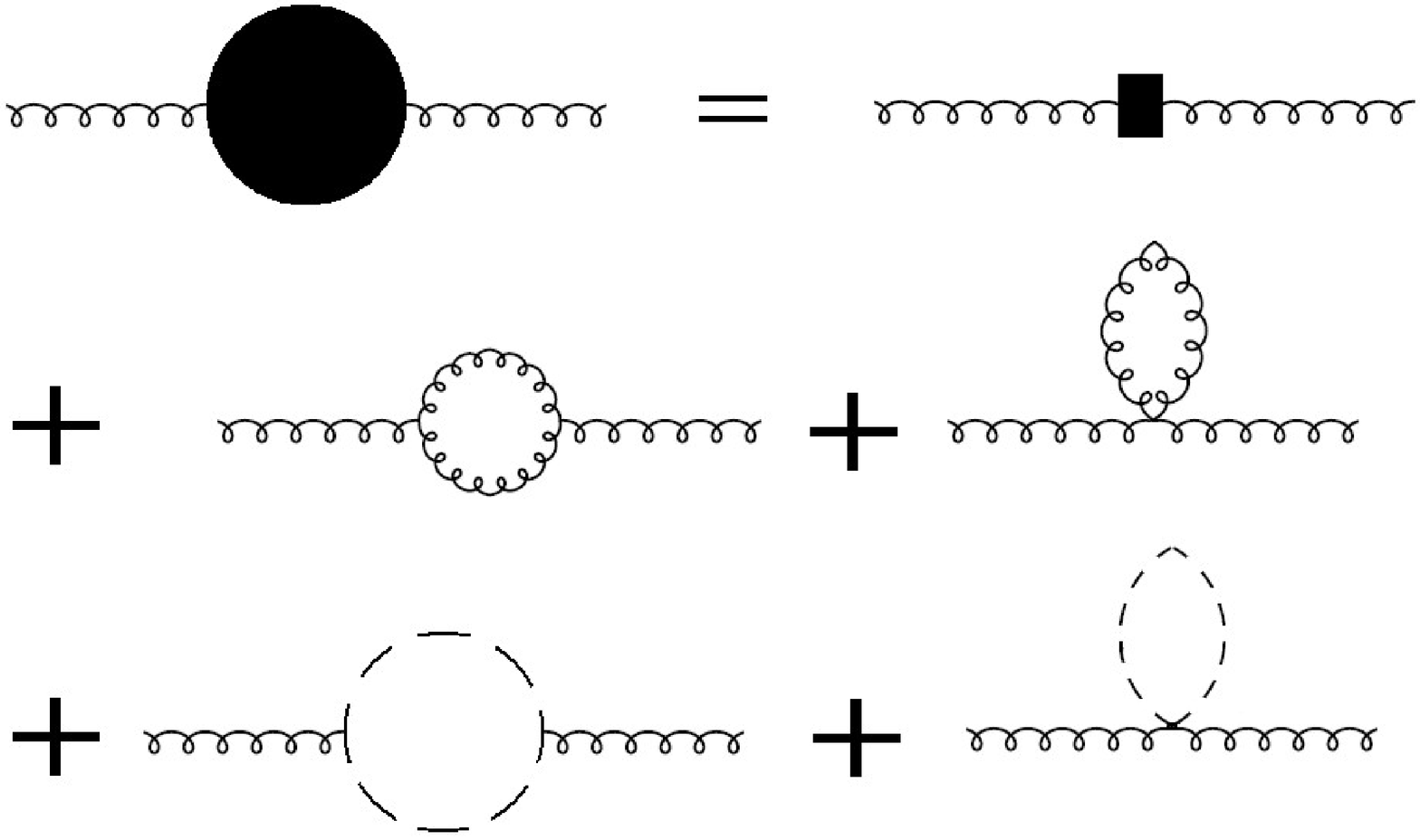}\\
{\footnotesize Fig. 5. Non-fermionic diagrams renormalizing the gluon
  propagator at 1 loop.}
\end{center}

\section{}

Tables I-IV provide a {\it per diagram} breakdown of our results, at a
given value of $m$ ($m=0.038$), in order to allow for potential
comparisons and cross checks. The total results for the coefficients
$h_2^{(j)}$, $h_{30}^{(j)}$, $h_{31}^{(j)}$, $h_{32}^{(j)}$ are listed
in Tables V-VIII, respectively, for a wide selection of $m$ values
which are used in the literature.
Given the smooth dependence of all these coefficients on $m$,
interpolations to other intermediate values of $m$ can be performed
with great accuracy.



\begin{table}[ht]
\begin{center}
\begin{minipage}{15cm}
\caption{Per-diagram contributions to $h^{(j)}_2$ . $m=0.038$ .
\label{m0038-2loop}}
\begin{tabular}{cr@{}lr@{}lr@{}l}
\multicolumn{1}{c}{diagram}&
\multicolumn{2}{c}{$h^{(0)}_2\cdot 10^4$} &
\multicolumn{2}{c}{$h^{(1)}_2\cdot 10^4$} &
\multicolumn{2}{c}{$h^{(2)}_2\cdot 10^4$} \\
\tableline \hline
37	&0&.932189(6) 		&0&.14684(1)  	&-5&.903340(3) \\
38	&-7&.442427121(2)	&0&		&0&\\
\end{tabular}
\end{minipage}
\end{center}
\end{table}


\begin{table}[ht]
\begin{center}
\begin{minipage}{15cm}
\caption{Per-diagram contributions to $h^{(j)}_{30}$ . $m=0.038$ .
\label{m0038-3loop0}}
\begin{tabular}{cr@{}lr@{}lr@{}lr@{}lr@{}l}
\multicolumn{1}{c}{diagram}&
\multicolumn{2}{c}{$h^{(0)}_{30}\cdot 10^4$} &
\multicolumn{2}{c}{$h^{(1)}_{30}\cdot 10^4$} &
\multicolumn{2}{c}{$h^{(2)}_{30}\cdot 10^4$} &
\multicolumn{2}{c}{$h^{(3)}_{30}\cdot 10^4$} &
\multicolumn{2}{c}{$h^{(4)}_{30}\cdot 10^4$} \\
\tableline \hline
39	&0&.19218007757(5)	&0&		&0&		&0&		&0& \\
41	&-0&.096284(2)		&-0&.047087(2)	&3&.176923(5)	&0&		&0& \\
42	&0&.00427805(2)		&0&		&-1&.303406(1)	&0&		&0& \\
43{+}53	&-1&.49693(1)		&0&.031601(2)	&-1&.343093(3)	&0&		&0& \\
45	&1&.574326(1)		&0&		&0&		&0&		&0& \\
46{+}56	&0&.102791(5)		&-0&.005620(1)	&0&.1231742(3)	&0&		&0& \\
47	&-0&.0006435(4)		&-0&.0369073(4)	&-0&.0637549(3)	&0&.00356648(1)	&0& \\
48	&0&			&-0&.0000025(1)	&-0&.0395366(6)	&0&		&0& \\
49	&0&.12905(6)		&-0&.01919(3)	&-0&.017117(8)	&-0&.0078566(7)	&0& \\
51{+}57	&-0&.37601(2)		&0&.023672(5)	&-0&.4837191(5)	&0&		&0& \\
54	&-0&.24831(4)		&-0&.3910(2)	&-0&.41414(8)	&0&		&0& \\
55{+}60	&0&.028358(1)		&-0&.0017428(5)	&0&.03556619(6)	&0&		&0& \\
59	&0&.012859(2)		&0&.01395(1)	&0&.07042(2)	&0&.05069(3)	&-0&.020929(8) \\
61{+}62	&0&.20558(1)		&-0&.011239(2)	&0&.2463485(6)	&0&		&0& \\
\end{tabular}
\end{minipage}
\end{center}
\end{table}


\begin{table}[ht]
\begin{center}
\begin{minipage}{15cm}
\caption{Per-diagram contributions to $h^{(j)}_{31}$ . $m=0.038$ .
\label{m0038-3loop1}}
\begin{tabular}{cr@{}lr@{}lr@{}lr@{}lr@{}l}
\multicolumn{1}{c}{diagram}&
\multicolumn{2}{c}{$h^{(0)}_{31}\cdot 10^4$} &
\multicolumn{2}{c}{$h^{(1)}_{31}\cdot 10^4$} &
\multicolumn{2}{c}{$h^{(2)}_{31}\cdot 10^4$} &
\multicolumn{2}{c}{$h^{(3)}_{31}\cdot 10^4$} &
\multicolumn{2}{c}{$h^{(4)}_{31}\cdot 10^4$} \\
\tableline \hline
39	&-0&.2882701164(1)	&0&		&0&		&0&		&0& \\
41	&0&.144427(2)		&0&.048077(2)	&-2&.951676(8)	&0&		&0& \\
42	&-0&.00855609(4)	&0&		&0&		&0&		&0& \\
43{+}53	&1&.62756(1)		&-0&.036704(2)	&1&.475838(3)	&0&		&0& \\
45	&-1&.574326(1)		&0&		&0&		&0&		&0& \\
47	&0&.0012870(5)		&0&.0073718(6)	&0&.0042472(4)	&0&		&0& \\
50	&-0&.0096(1)		&-0&.02767(1)	&0&.01081(4)	&0&.000361(3)	&-0&.007025(2) \\
54	&0&.24831(5)		&0&.3910(2)	&0&.41414(8)	&0&		&0& \\
59	&-0&.012859(2)		&-0&.01395(2)	&-0&.07042(2)	&-0&.05069(3)	&0&.020929(8) \\
\end{tabular}
\end{minipage}
\end{center}
\end{table}


\begin{table}[ht]
\begin{center}
\begin{minipage}{15cm}
\caption{Per-diagram contributions to $h^{(j)}_{32}$ . $m=0.038$ .
\label{m0038-3loop2}}
\begin{tabular}{cr@{}lr@{}lr@{}lr@{}lr@{}l}
\multicolumn{1}{c}{diagram}&
\multicolumn{2}{c}{$h^{(0)}_{32}\cdot 10^6$} &
\multicolumn{2}{c}{$h^{(1)}_{32}\cdot 10^6$} &
\multicolumn{2}{c}{$h^{(2)}_{32}\cdot 10^6$} &
\multicolumn{2}{c}{$h^{(3)}_{32}\cdot 10^6$} &
\multicolumn{2}{c}{$h^{(4)}_{32}\cdot 10^6$} \\
\tableline \hline
44{+}52{+}58	&3&.65818(5) 	&-0&.24532(8)    &7&.62942(7)	&-0&.41017(3)	&5&.742740(1) \\
\end{tabular}
\end{minipage}
\end{center}
\end{table}



\begin{table}[ht]
\begin{center}
\begin{minipage}{15cm}
\caption{Total values of $h^{(j)}_2$, for various masses.
\label{total2loop}}
\begin{tabular}{r@{}lr@{}lr@{}lr@{}l}
\multicolumn{2}{c}{$m$}&
\multicolumn{2}{c}{$h^{(0)}_2\cdot 10^3$} &
\multicolumn{2}{c}{$h^{(1)}_2\cdot 10^3$} &
\multicolumn{2}{c}{$h^{(2)}_2\cdot 10^3$} \\
\tableline \hline
-1&.014925	&-2&.4083635(4)	&0&.395934967(3)	&-0&.71409763(8) \\
-0&.9512196	&-2&.2738528(3)	&0&.359466059(5)	&-0&.70919125(2) \\
-0&.8749999	&-2&.1173032(2)	&0&.318089891(2)	&-0&.70272042(1) \\
-0&.8253968	&-2&.0179859(2)	&0&.292493808(2)	&-0&.698193815(6) \\
-0&.7948719	&-1&.9578750(2)	&0&.277266040(2)	&-0&.695294722(3) \\
-0&.5181059	&-1&.4485862(6)	&0&.15739842(6)		&-0&.66565598(2) \\
-0&.423462	&-1&.2897989(2)	&0&.1238792(6)		&-0&.6543190(2) \\
-0&.4028777	&-1&.2563603(3)	&0&.1170884(9)		&-0&.6517790(3) \\
-0&.3140433	&-1&.1167334(5)	&0&.089818(2)		&-0&.6405147(6) \\
-0&.301775	&-1&.0980649(6)	&0&.086312(1)		&-0&.6389201(5) \\
-0&.2962964	&-1&.0897779(5)	&0&.084766(2)		&-0&.6382049(7) \\
-0&.2852897	&-1&.0732218(2)	&0&.081700(5)		&-0&.636761(3) \\
-0&.2769916	&-1&.0608234(1)	&0&.079422(6)		&-0&.635665(4) \\
-0&.2686568	&-1&.0484419(4)	&0&.077161(3)		&-0&.634562(4) \\
-0&.1482168	&-0&.8779913(3)	&0&.04774(1)		&-0&.618118(2) \\
0&.		&-0&.6929202(1)	&0&.02010061(2)		&-0&.59630769(1) \\
0&.038		&-0&.6510238(6)	&0&.014684(1)		&-0&.5903340(3) \\
0&.072		&-0&.615948(1)	&0&.010488(4)		&-0&.584812(2) \\
0&.0927		&-0&.5956994(5)	&0&.008214(2)		&-0&.5813703(7) \\
\end{tabular}
\end{minipage}
%
\vskip 0.2cm
%
\begin{minipage}{15cm}
\caption{Total values of $h^{(j)}_{30}$, for various masses.
\label{total3loop0}}
\begin{tabular}{r@{}lr@{}lr@{}lr@{}lr@{}lr@{}l}
\multicolumn{2}{c}{$m$}&
\multicolumn{2}{c}{$h^{(0)}_{30}\cdot 10^4$} &
\multicolumn{2}{c}{$h^{(1)}_{30}\cdot 10^4$} &
\multicolumn{2}{c}{$h^{(2)}_{30}\cdot 10^4$} &
\multicolumn{2}{c}{$h^{(3)}_{30}\cdot 10^4$} &
\multicolumn{2}{c}{$h^{(4)}_{30}\cdot 10^4$} \\
\tableline \hline
-1&.014925	&-3&.61593(9)	&-0&.5205(4)	&-0&.16938(2)	&0&.18044(3)	&-0&.01301953(3) \\
-0&.9512196	&-3&.30186(3)	&-0&.5660(4)	&-0&.16042(2)	&0&.17452(3)	&-0&.01362776(1) \\
-0&.8749999	&-2&.93686(3)	&-0&.6119(4)	&-0&.15115(2)	&0&.16699(4)	&-0&.014206930(3) \\
-0&.8253968	&-2&.70587(4)	&-0&.6368(5)	&-0&.14582(2)	&0&.16186(4)	&-0&.014512957(7) \\
-0&.7948719	&-2&.56636(4)	&-0&.6503(5)	&-0&.14275(1)	&0&.15863(4)	&-0&.01467848(1) \\
-0&.5181059	&-1&.4032(2)	&-0&.7097(3)	&-0&.1185(1)	&0&.126761(7)	&-0&.01575561(2) \\
-0&.423462	&-1&.0538(3)	&-0&.7039(1)	&-0&.1097(1)	&0&.11495(2)	&-0&.016111(1) \\
-0&.4028777	&-0&.9817(2)	&-0&.7010(1)	&-0&.10756(6)	&0&.11233(2)	&-0&.016198(2) \\
-0&.3140433	&-0&.6879(1)	&-0&.6810(1)	&-0&.0971(1)	&0&.100645(8)	&-0&.016638(4) \\
-0&.301775	&-0&.6497(1)	&-0&.6770(3)	&-0&.0956(1)	&0&.09905(3)	&-0&.016708(4) \\
-0&.2962964	&-0&.6329(1)	&-0&.6751(2)	&-0&.0948(1)	&0&.09828(5)	&-0&.016735(8) \\
-0&.2852897	&-0&.5994(1)	&-0&.6719(4)	&-0&.0934(1)	&0&.09675(5)	&-0&.01681(1) \\
-0&.2769916	&-0&.5746(1)	&-0&.6686(5)	&-0&.0921(2)	&0&.09565(5)	&-0&.016868(9) \\
-0&.2686568	&-0&.54989(8)	&-0&.6653(5)	&-0&.0907(2)	&0&.09458(3)	&-0&.016911(8) \\
-0&.1482168	&-0&.23130(6)	&-0&.6101(4)	&-0&.0704(2)	&0&.07775(9)	&-0&.01785(2) \\
0&.		&0&.01867(6)	&-0&.49191(1)	&-0&.02955(1)	&0&.053914(1)	&-0&.01998581(2) \\
0&.038		&0&.03126(8)	&-0&.4436(2)	&-0&.01234(8)	&0&.04640(3)	&-0&.020929(8) \\
0&.072		&0&.03095(9)	&-0&.3997(2)	&0&.00431(7)	&0&.03982(2)	&-0&.021807(5) \\
0&.0927		&0&.02828(7)	&-0&.3744(3)	&0&.0144(1)	&0&.03610(3)	&-0&.022300(6) \\
\end{tabular}
\end{minipage}
\end{center}
\end{table}


\begin{table}[ht]
\begin{center}
\begin{minipage}{15cm}
\caption{Total values of $h^{(j)}_{31}$, for various masses.
\label{total3loop1}}
\begin{tabular}{r@{}lr@{}lr@{}lr@{}lr@{}lr@{}l}
\multicolumn{2}{c}{$m$}&
\multicolumn{2}{c}{$h^{(0)}_{31}\cdot 10^4$} &
\multicolumn{2}{c}{$h^{(1)}_{31}\cdot 10^4$} &
\multicolumn{2}{c}{$h^{(2)}_{31}\cdot 10^4$} &
\multicolumn{2}{c}{$h^{(3)}_{31}\cdot 10^4$} &
\multicolumn{2}{c}{$h^{(4)}_{31}\cdot 10^4$} \\
\tableline \hline
-1&.014925	&4&.2718(2)	&0&.61383(4)	&-1&.1331(2)	&-0&.18127(4)	&0&.002886(2) \\
-0&.9512196	&3&.9078(2)	&0&.636745(4)	&-1&.1391(2)	&-0&.17467(4)	&0&.003690(2) \\
-0&.8749999	&3&.4867(2)	&0&.65829(6)	&-1&.1434(2)	&-0&.16655(3)	&0&.004505(1) \\
-0&.8253968	&3&.2212(2)	&0&.6689(1)	&-1&.1447(2)	&-0&.16119(3)	&0&.004966(1) \\
-0&.7948719	&3&.0614(2)	&0&.6741(1)	&-1&.1450(2)	&-0&.15783(3)	&0&.005226(1) \\
-0&.5181059	&1&.7413(2)	&0&.6750(2)	&-1&.1343(1)	&-0&.12639(1)	&0&.007141(2) \\
-0&.423462	&1&.3494(2)	&0&.6559(1)	&-1&.12732(6)	&-0&.11512(2)	&0&.007770(1) \\
-0&.4028777	&1&.2686(2)	&0&.6503(1)	&-1&.12575(6)	&-0&.11264(1)	&0&.007917(2) \\
-0&.3140433	&0&.9409(2)	&0&.6207(1)	&-1&.11893(5)	&-0&.10162(2)	&0&.008611(6) \\
-0&.301775	&0&.8984(2)	&0&.6156(3)	&-1&.1180(1)	&-0&.10013(4)	&0&.008715(5) \\
-0&.2962964	&0&.8797(2)	&0&.6132(1)	&-1&.1176(1)	&-0&.09941(6)	&0&.008759(8) \\
-0&.2852897	&0&.8424(2)	&0&.6090(3)	&-1&.1168(1)	&-0&.09798(5)	&0&.00887(1) \\
-0&.2769916	&0&.8147(2)	&0&.6050(4)	&-1&.1163(2)	&-0&.09695(6)	&0&.00895(1) \\
-0&.2686568	&0&.7872(2)	&0&.6011(4)	&-1&.1159(2)	&-0&.09594(3)	&0&.009012(8) \\
-0&.1482168	&0&.4324(2)	&0&.5382(4)	&-1&.1090(2)	&-0&.0801(1)	&0&.01029(2) \\
0&.		&0&.1479(2)	&0&.41633(5)	&-1&.11100(1)	&-0&.057537(3)	&0&.01284418(4) \\
0&.038		&0&.1279(1)	&0&.3681(2)	&-1&.1171(1)	&-0&.05033(3)	&0&.013904(9) \\
0&.072		&0&.1221(1)	&0&.3248(2)	&-1&.1232(1)	&-0&.04400(2)	&0&.014891(5) \\
0&.0927		&0&.1211(1)	&0&.3000(3)	&-1&.1268(1)	&-0&.04044(3)	&0&.015451(6) \\
\end{tabular}
\end{minipage}
%
\vskip 0.2cm
%
\begin{minipage}{15cm}
\caption{Total values of $h^{(j)}_{32}$, for various masses.
\label{total3loop2}}
\begin{tabular}{r@{}lr@{}lr@{}lr@{}lr@{}lr@{}l}
\multicolumn{2}{c}{$m$}&
\multicolumn{2}{c}{$h^{(0)}_{32}\cdot 10^4$} &
\multicolumn{2}{c}{$h^{(1)}_{32}\cdot 10^4$} &
\multicolumn{2}{c}{$h^{(2)}_{32}\cdot 10^4$} &
\multicolumn{2}{c}{$h^{(3)}_{32}\cdot 10^4$} &
\multicolumn{2}{c}{$h^{(4)}_{32}\cdot 10^4$} \\
\tableline \hline
-1&.014925	&0&.49011557(4)	&-0&.19966299(3)&0&.36828835(3)	&-0&.11015360(1)&0&.08360193(2) \\
-0&.9512196	&0&.43836868(2)	&-0&.17353447(1)&0&.343436025(6)&-0&.10025167(1)&0&.082274351(7) \\
-0&.8749999	&0&.38167686(2)	&-0&.14550571(1)&0&.31500822(1)	&-0&.088919069(7)&0&.080600027(9) \\
-0&.8253968	&0&.34767753(1)	&-0&.129038814(4)&0&.297262047(4)&-0&.081862310(7)&0&.079469961(8) \\
-0&.7948719	&0&.327840761(5)&-0&.119564143(4)&0&.286632961(4)&-0&.077648919(7)&0&.078761444(7) \\
-0&.5181059	&0&.18222240(4)	&-0&.0539075(2)	&0&.20003143(6)	&-0&.04416150(5)&0&.072038437(4) \\
-0&.423462	&0&.1450777(2)	&-0&.0386899(4)	&0&.1742885(2)	&-0&.0347205(4)	&0&.06966149(6) \\
-0&.4028777	&0&.1377609(1)	&-0&.0357991(4)	&0&.1689417(2)	&-0&.0328053(6)	&0&.0691407(1) \\
-0&.3140433	&0&.1091234(3)	&-0&.0249053(4)	&0&.1468864(3)	&-0&.025111(1)	&0&.0668770(3) \\
-0&.301775	&0&.1055301(5)	&-0&.023593(2)	&0&.1439703(7)	&-0&.024122(3)	&0&.0665614(9) \\
-0&.2962964	&0&.1039530(7)	&-0&.023021(2)	&0&.1426783(9)	&-0&.023686(3)	&0&.0664209(8) \\
-0&.2852897	&0&.1008352(9)	&-0&.021900(2)	&0&.1401015(9)	&-0&.022821(1)	&0&.0661380(3) \\
-0&.2769916	&0&.0985290(7)	&-0&.021075(1)	&0&.1381755(8)	&-0&.022178(4)	&0&.0659243(2) \\
-0&.2686568	&0&.0962507(4)	&-0&.020271(6)	&0&.1362553(5)	&-0&.021542(4)	&0&.0657091(3) \\
-0&.1482168	&0&.0674197(9)	&-0&.010680(4)	&0&.1101470(8)	&-0&.013260(4)	&0&.062561(1) \\
0&.		&0&.04159747(5)	&-0&.00356316(1)&0&.08245984(4)	&-0&.005574256(7)&0&.058518838(6) \\
0&.038		&0&.0365818(5)	&-0&.0024532(8)	&0&.0762942(7)	&-0&.0041017(3)	&0&.05742740(1) \\
0&.072		&0&.0326353(4)	&-0&.001675(1)	&0&.071174(1)	&-0&.0029808(5)	&0&.0564209(3) \\
0&.0927		&0&.0304618(8)	&-0&.0012878(2)	&0&.0682377(9)	&-0&.002377(2)	&0&.0557953(1) \\
\end{tabular}
\end{minipage}
\end{center}
\end{table}

\end{document}